\documentclass[a4paper]{jpconf}

\usepackage{graphicx}
\usepackage{hyperref}

\begin{document}
\title{The LIGO Open Science Center}

\author{Michele Vallisneri$^{1}$, Jonah Kanner$^{2}$, Roy Williams$^{2}$, Alan Weinstein$^{2}$ and Branson Stephens$^{3}$}

\address{${}^1$ Jet Propulsion Laboratory, California Institute of Technology, Pasadena CA 91109, USA}
\address{${}^2$ LIGO -- California Institute of Technology, Pasadena CA 91125, USA}
\address{${}^3$ University of Wisconsin -- Milwaukee, Milwaukee WI 53201, USA}

\ead{michele.vallisneri@jpl.nasa.gov}

\begin{abstract}
The LIGO Open Science Center (LOSC) fulfills LIGO's commitment to release, archive, and serve LIGO data in a broadly accessible way to the scientific community and to the public, and to provide the information and tools necessary to understand and use the data.
In August 2014, the LOSC published the full dataset from Initial LIGO's ``S5'' run at design sensitivity, the first such large-scale release and a valuable testbed to explore the use of LIGO data by non-LIGO researchers and by the public, and to help teach gravitational-wave data analysis to students across the world. In addition to serving the S5 data, the LOSC web portal (losc.ligo.org) now offers documentation, data-location and data-quality queries, tutorials and example code, and more.
We review the mission and plans of the LOSC, focusing on the S5 data release.
\end{abstract}

\section{Introduction}


The emerging \emph{Open Science} movement holds that the unprecedented ease of sharing information and collaborating remotely on the internet will foster a new era of greatly accelerated and amplified discovery, \emph{if} data, publications, and results are made available openly and broadly \cite{nielsen}.
The 2010 ``Panton principles'' open-data manifesto \cite{panton} characterizes science as based on ``building on, reusing and openly criticising the published body of scientific knowledge,'' and argues that ``for science to effectively function, and for society to reap the full benefits from scientific endeavours, it is crucial that science data be made open.''
The U.S.\ government accepted these arguments, and in 2013 it issued a general call to its agencies to make ``information resources accessible, discoverable, and usable by the public'' to ``help fuel entrepreneurship, innovation, and scientific discovery'' \cite{OMB}. This memorandum characterizes \emph{open data} as ``fully discoverable and usable by end users,'' which it spells out further as public, accessible, described, reusable, complete, timely, and managed in post-release.

LIGO, the Laser Interferometer Gravitational-wave Observatory \cite{2009RPPh...72g6901A}, is a facility dedicated to measuring cosmic gravitational waves (GWs); it is funded by the U.S.\ National Science Foundation (NSF) and operated by the California Institute of Technology and the Massachusetts Institute of Technology. LIGO has been collecting science data since 2002. Analyzing these data is one of the responsibilities of the LIGO Scientific Collaboration (LSC), which now comprises $\sim$ 90 institutions and 900 members worldwide. 
While the LSC is open to any contributor from the global community, any astrophysical publications based on LIGO data must currently be approved by the collaboration as a whole, and then published with a full list of authors.
In 2008, the NSF asked LIGO and the LSC to devise a roadmap to open the LIGO data to professional researchers outside the LSC and to a broader public that includes amateur scientists, students, and more. 

There are many good reasons why the LIGO data should be open.
First, since LIGO is publicly funded, \emph{the public owns the data}. 
Next and most important, opening the LIGO data will \emph{maximize discovery}. It will benefit the LSC data-analysis efforts, enabling reproducibility, getting more eyes on LIGO data to recognize potential problems and solutions more rapidly, and improving the communication of LSC results to colleagues. It will also benefit collaborations with other GW and non-GW projects, facilitating electromagnetic followups and other multimessenger astronomy, and harnessing new perspectives and insights to advance LSC science. Last, open data access will serve to interest and engage a \emph{much wider community}, bringing the excitement of GW science and multimessenger astronomy to students, amateur and professional astronomers and scientists in related fields.

LIGO's data-release roadmap \cite{dataplan}, which later became part of LIGO's data management plan \cite{datamanagement}, identifies two phases to the release.
During the early \emph{discovery phase}, which will start as the upgraded Advanced LIGO detectors \cite{2010CQGra..27h4006H} begin to collect science data (as early as 2015), the understanding of the LIGO measurements is evolving as the instrument is tuned toward design sensitivity, and detections are rare. Use of the full data is restricted to the LSC, although early \emph{transient alerts} of candidate GW events are broadcast to astronomers who have signed on to search for electromagnetic counterparts. The LSC releases to the public segments of GW strain data around validated discoveries, or astrophysically significant nondetections (such a release happened already for short-hard GRB 070201 after LIGO was able to exclude that the burst originated from a neutron-star binary coalescence in M31 \cite{2008ApJ...681.1419A}).

The \emph{observational phase} begins after one of three conditions occurs: LIGO has obtained plentiful detections; it has probed a space-time volume of $3 \times 10^7 \, \mathrm{Mpc}^3 \, \mathrm{yr}$; or 3.5 years have elapsed since its formal acceptance.
In this phase, the sensitivity of the observatory and the mature understanding of the data allow the exploration of the astrophysical content of the measured GWs.
The entire dataset is released to the public in six-month segments with 24-month latency, as necessary to correct the data for instrumental and environmental issues \cite{datamanagement}. In this phase also transient alerts are issued publicly.

\section{The LIGO Open Science Center}

Archiving and serving scientific data requires a dedicated service, characterized formally by the International Organization for Standardization (ISO) as an \emph{Open Archival Information System} (OAIS, \cite{oais}). An OAIS includes elements that perform data ingestion, storage, annotation, preservation, operations, and access.
Such a system exists already in the form of the \emph{LIGO Data System}, which has archived one petabyte of LIGO data (1\% of which in the GW channel, the rest in a myriad of environmental and instrument-monitoring channels) and has made it available within seconds of data taking to hundreds of LSC scientists.

However, the functions of an OAIS need to be fulfilled with respect to the needs and competences of one or more \emph{designated communities} of data consumers. Consider for instance data preservation, which extends to the \emph{bits} as well as the \emph{meaning} of the data: achieving this goal for LSC members, who can rely the support of their peers and on the tacit knowledge obtained in their professional training, is clearly a very different task than doing so for non-LSC scientists and for the general public.
To extend the LIGO Data System to these new designated communities, LIGO created the LIGO Open Science Center (LOSC), defined \cite{datamanagement} as ``a fabric to enable robust scientific discourse about the LIGO data for somebody who has never met any LSC member.''
Who may be the users of the LOSC data? Scientists with interests in GWs, relativity, astrophysics, and multimessenger astronomy; undergraduate and graduate students in the sciences; even K-12 STEM students and their teachers.

The LOSC team (comprising the named authors of this article) began work in 2012, focusing on an initial set of milestones for documenting, distributing, and curating the LIGO data:
\begin{itemize}
\item releasing the full GW strain data from the 2005--2007 ``S5'' science run, which achieved design sensitivity for Initial LIGO;
\item validating the GW strain data and annotating it with easily interpreted data-quality information and with information about the simulated GW signals present in the data (a.k.a.\ \emph{hardware injections});
\item setting up the hardware and software for a LOSC web portal, \url{losc.ligo.org}, offering access to open data as well as additional services; 
\item providing a database service for the available data and its quality;
\item producing a suite of tutorials and example codes to acquire, load, manipulate, filter, and plot the data. 
\end{itemize}

As we write this article, we have reached all these milestones.
Most notably, in August 2014 the LOSC released an open dataset for LIGO's flagship fifth science run S5 (which took place between November 2005 and October 2007), the first in which LIGO performed at design sensitivity. The LSC, in partnership with the Virgo Collaboration \cite{2012JInst...7.3012A}, searched the data for a variety of sources, and published a number of papers limiting the rates of GW-emitting events in the local universe (\cite{2009PhRvD..79l2001A,2009PhRvD..80j2001A,2009PhRvL.102k1102A} and several more, see \cite{lscpp} for a complete list of LSC publications).
The S5 dataset includes data from the three LIGO instruments (the 4-km H1 and 2-km H2 in Hanford, WA, and the 4-km L1 in Livingston, LA), which collected 537, 546, and 457 days of science data, respectively.

We intend for the S5 dataset to serve as a foundation for future data releases (including the ``S6'' data, which is currently under internal review); we anticipate that it will create a realistic testbed to engage the scientific community, the general public and the LSC itself in a discussion about the most useful treatments and representations of the LIGO data products; and we trust that it will provide a valuable tool to teach GW data analysis to students at different levels. We plan to continue managing and supporting access to the S5 dataset as we prepare for further releases.

In the remainder of this proceeding, we describe the nature and technical details of the data archived by the LOSC, and the main services provided by the LOSC portal. 
 
\section{The LOSC data archive}

The LIGO detectors generate thousands of channels of data ($\sim$ 6,600 for Initial LIGO \cite{2009RPPh...72g6901A}) corresponding to many different types of measurements and instrument monitors. The LOSC offers three simpler data products, which (we expect) will fulfill the scientific and pedagogical requirements of its designated communities. These products are the 4,096-Hz time series of the calibrated GW observable (known as \emph{strain}, or $h(t)$); a 1-Hz time series that characterizes the \emph{quality} of the strain data, by summarizing the data cuts and vetoes made by the LSC scientists who searched the S5 data for GWs; and another 1-Hz time series that flags the simulated GWs \emph{injections} that were inserted in the data by actuating the LIGO test masses, to test search pipelines and characterize their performance.

The three LOSC time series are packaged together in files that each cover 4,096 contiguous seconds (an arbitrary but convenient choice). In designing and creating these files, we followed the principles discussed in the LIGO data-management plan \cite{datamanagement}, requiring LOSC data to be:
\begin{itemize}
\item \emph{Interoperable}: The LOSC files are available in the broadly used and supported HDF5 format \cite{hdf5}, and also in the \emph{frame} format \cite{frame} used internally by the GW experimenters; free and open software tools to read both formats are available or linked on the LOSC portal. 
\item \emph{Self-describing}: The LOSC files contain metadata describing their content, as well as ``keys'' to interpret the data-quality and injection time series.
\item \emph{Published online}: The LOSC files can be located and downloaded easily from the LOSC portal \emph{by click} (i.e., through a web browser) and \emph{by code} (i.e., by automated scripts that use a well-defined interface).
\item \emph{Contextualized}: The LOSC portal offers a ``getting started'' page about the files and their content, example code and tutorials, extensive documentation and references.
\item \emph{Supported}: The LOSC is organizing workshops and webinars about the data, will provide ``help desk'' functions as resources allow, and will continue to refine the datasets and correct problems if any are found.
\end{itemize}

This section continues with more details about the three time series contained in the LOSC data files.

\subsection{GW strain}
This time series represents the dimensionless GW strain $h(t)$, conceptually the fractional change induced by GWs in the LIGO arm length, as measured with laser interferometry. The $h(t)$ data were created by LSC analysts by \emph{calibrating} \cite{2010NIMPA.624..223A} the LIGO interferometric measurements with a variety of techniques to infer actual arm-length changes from optical observables.
These data record the GWs impinging on the detectors plus instrumental (strain-equivalent) noise; since no GW events were found in the S5 run, it is fair to say that the S5 $h(t)$ data are, to our knowledge, essentially all noise.

We downsample $h(t)$ to 4 kHz from the original 16 kHz to reduce the size of LOSC files while (arguably) losing no science content, since the higher frequencies are dominated by optical noise.
Thus, a 4,096-second file contains 16,777,216 GW-strain samples from a single detector, represented as floating-point ``doubles'' (eight bytes each). If samples are missing because the detector was not taking science data at that time, they are represented as ``NaN'' (not a number) values. Each file has a starting GPS time\footnote{Time as obtained from an accurate GPS receiver, given as the number of seconds since the January 6th, 1980 epoch, not including leap seconds.} corresponding to an integer multiple of 4,096. Such a file takes up to 120 MB, allowing for compression; the whole S5 dataset (data from three detectors over two years, with roughly 65\% duty cycle)
has a size of $3 \times 2 \times 365 \times 24 \times 3600 \times 4096 \times 0.65 \times 8 \,\mathrm{bytes} = $ almost 4 TB.

\subsection{Data quality}

The LIGO instruments did not continuously collect data, since operation was frequently interrupted by commissioning work, seismic events, intermittent construction or forest logging nearby, and other causes. As a result, the S5 dataset contains many data gaps, $\sim 30\%$ of the total data-taking period. Most LSC searches would therefore begin from a \emph{segment list} indicating intervals of existing science data, usually aligned to whole seconds.
To capture this information, we label every second of usable science data with the marker \texttt{DATA} (and conversely, all the GW-strain samples in intervals not marked by \texttt{DATA} are NaNs in the LOSC files).

In addition to this minimal characterization, the GW searches performed by LSC scientists adopted further judgments about the quality of science data, ranking every second of observation in one of four \emph{categories}. They performed the ranking by developing hundreds of \emph{vetoes} that identify instrumental and environmental conditions (ranging from passing trains, saturated photodiodes, and powerline fluctuations to high winds, elevated seismic noise, and magnetometer spikes) found to be correlated to glitches and increased noise in the GW-strain channel. They then assigned each veto to a category, with slightly different assignments for different astrophysical searches (such as the search for GW signals from low-mass binary inspirals, known as \texttt{CBCLOW} \cite{2009PhRvD..79l2001A}, or the all-sky search for GW bursts, known as \texttt{BURST} \cite{2009PhRvD..80j2001A}). Overall, the meaning of the four categories can be summarized as follows:
\begin{itemize}
\item Data passing \texttt{CAT1} vetoes is free of severe problems, so the detector is considered to be in observation mode.
\item Data passing \texttt{CAT2} (in addition to \texttt{CAT1}) vetoes is also free of problems resulting from well understood couplings between the GW-strain and instrumental channels. In LIGO searches, data passing \texttt{CAT2} is typically used to identify GW event candidates.
\item Data passing \texttt{CAT3} (in addition to \texttt{CAT1} and \texttt{CAT2}) vetoes is also free of problems resulting from less well established couplings between the GW and instrumental channels. In LIGO searches, data passing \texttt{CAT3} is typically used to derive GW upper limits. 
\item Data passing \texttt{CAT4} (in addition to \texttt{CAT1}--\texttt{CAT3}) vetoes is also free of problems resulting from poorly established couplings between the GW and instrumental channels.
\end{itemize}

The LOSC data-quality time series is given at 1 Hz cadence, and it consists of short \emph{bitmasks} where setting an individual bit denotes that for that whole second (corresponding to 4,096 GW-strain samples) the LIGO data passed ``\texttt{CATn}'' vetoes according to the veto definitions of a certain search (for S5, one of \texttt{CBCHIGH} \cite{2010PhRvD..82j2001A}, \texttt{CBCLOW} \cite{2009PhRvD..79l2001A}, \texttt{BURST} \cite{2009PhRvD..80j2001A}, \texttt{CONTINUOUS} \cite{2009PhRvL.102k1102A}, and \texttt{STOCHASTIC} \cite{2009Natur.460..990A}). In addition, setting bit 0 corresponds to data possessing the \texttt{DATA} marker discussed above. The meaning of the data-quality bits is described in the metadata contained in each LOSC file.

Using this data-quality time series, users of the LOSC data may construct segment lists to analyze according to the most fitting veto definitions. For example, scientists mounting a search for short-duration GWs may use the segments of data for which the \texttt{BURST\_CAT2} bit is set, while those looking for longer transients may use \texttt{CBCHIGH\_CAT2}, or perhaps \texttt{CBCHIGH\_CAT3} if setting upper limits. As we describe below, the LOSC website contains example code to form these segment lists.

\subsection{Injections}

Thousands of simulated GW signals, known as \emph{hardware injections} (see, e.g., \cite{2004CQGra..21S.797B}), are routinely embedded in the LIGO strain measurements by actuating the LIGO test masses at scheduled (and carefully catalogued!) times. In S5, such injections were performed by the four LIGO search groups (those looking for compact binary coalescences, burst signals, continuous waves, and stochastic signals \cite{2009PhRvD..79l2001A,2009PhRvD..80j2001A,2009PhRvL.102k1102A,2009Natur.460..990A}), to test the efficiency and characterize the statistics of their search pipelines, and to validate the calibration of the detector response to GWs.
Search pipelines and detection protocols were further validated with a small number of \emph{blind} injections that were not disclosed to data analysts until the searches had been completed (see, e.g., \cite{bigdog}).
Binary and burst injections are short transients (several seconds to several minutes); continuous injections are longer-lived (potentially hours), but almost monochromatic, so they do not disturb the detection of transient signals occurring alongside them; by contrast, stochastic injections are both long-lived and broadband, so the loudest ones may invalidate, for the purpose of other searches, the stretches of data where they occur.

The LOSC data files include a 1-Hz bitmask timeline that identifies the seconds during which a hardware injection was active, as well as the type of the injection (when recorded in the LSC segment database). Thus, users of the LOSC data can build segment lists that exclude injection times, or focus on those very times to develop and test their search algorithms. In addition, separate listings of the times and parameters of all S5 hardware injections are available on the LSC portal.

\section{LOSC data services}

In addition to serving the HDF and frame files that contain the LOSC GW-strain, data-quality, and injection time series, the LOSC portal \url{losc.ligo.org} provides a variety of services and contextual information. We describe some highlights in this section.

\subsection{Archive database}

The LOSC portal offers access to a catalog of the LOSC files, which reproduces their basic metadata and a set of useful statistics: the minimum, maximum, standard deviation, and band-limited root-mean-square of the GW strain, as well as the number of seconds for which each data-quality and injection bit is set. The catalog can be queried to display the files that contain data between arbitrary GPS times, either as a table (color-coded by data quality and by the presence of injections), or in textual form, or in machine-readable JSON format \cite{json}.

\subsection{Timeline}

\begin{figure}
\begin{center}
\includegraphics[width=0.9\textwidth]{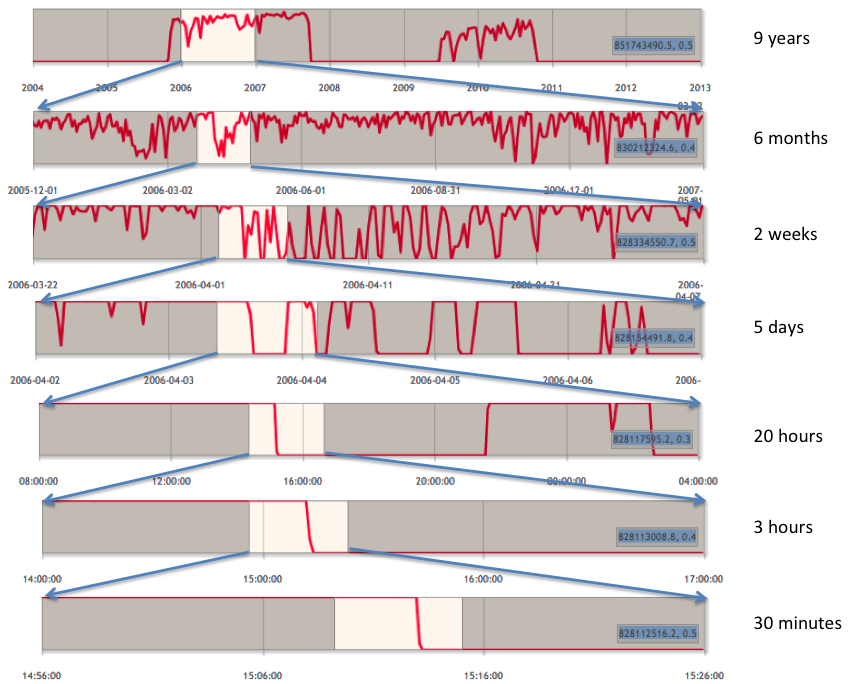}
\end{center} \vspace{-12pt}
\caption{The LOSC Timeline tool offers instantaneous access to \emph{duty-cycle} information for data-quality and injection bits over intervals of minutes to years.\label{fig:timeline}}
\end{figure}

The LOSC Timeline tool presents a visual representation of the \emph{duty cycle} of each data-quality (or injection) bit across a prescribed time interval: if the interval is short enough that each second can be shown onscreen, then the timeline for each bit is always either 0 or 1; for longer intervals, Timeline shows the average of the bit over nonoverlapping $2^n$-second subintervals. The averages are cached at various levels to offer immediate access to data-quality or injection history over hours, weeks, years, and decades (see Fig.\ \ref{fig:timeline}).
Multiple bits can be explored together interactively, or the underlying information can be fetched in machine-readable JSON format or text-based segment lists.


In addition to all data-quality and injection bits for the S5 run, the LOSC Timeline tool currently visualizes science-mode duty cycles for LIGO's international partners: the German--British GEO600 \cite{2010CQGra..27h4003G} and the French--Italian Virgo \cite{2012JInst...7.3012A}. Thus from a single web page we may explore the operation of five detectors, in three countries, over 12 years.

\subsection{MySources}

The LOSC MySources tool (\url{losc.ligo.org/mysources}) allows users to find out which GW detectors (among the three LIGO detectors, GEO, and Virgo) were taking data at the times of a user-provided collection of timed events, such as gamma-ray bursts, magnetar flares, or high-energy-particle measurements. The tool accepts the times (either UTC or GPS), event names, and (if desired) the sky positions of the events, and returns a listing of the observatories that have science data for those times, as well as the antenna patterns that quantify detector geometric response to a source at a given sky position.

\subsection{Tutorials and examples}

The LOSC portal also includes a set of tutorials and example codes (mostly in Python, but also in MATLAB and C) to learn how to download, parse, and manipulate the LOSC data files. 
The intention is to provide not a comprehensive data-analysis suite, which already exists in the public domain as the comprehensive (but challenging) LIGO Algorithm Library \cite{lal}, but rather small and transparent samples that a beginner can use to build up to more sophisticated tools, or just to explore and learn.
The tutorials cover software-environment prerequisites; searching for and downloading the files; inspecting them; working with the data-quality bitmasks to build segment lists; plotting time series, spectra, and spectrograms; identifying injections; and searching for (simple) chirp-like signals. A final tutorial explains how to use web services to automate the search of open data, how to navigate the metadata and catalogs, and how to download the entire S5 dataset file-by-file in a robust way.

\section{Summary}

The creation of the LOSC is an important step in aligning the GW community with the broad social movement towards open science, and in supporting the continuing trajectory of the international network of GW observatories toward a model of shared data analysis.
The LOSC provides public access to LIGO data products and to value-added documentation, tutorials, and (free and open) software, with the aim of making the LIGO data truly comprehensible and usable by a broad audience. Significant resources have been committed to ensuring that the data are presented accurately, and that the contextual information is sufficient to allow reuse by scientists, students, and others inside and outside the LIGO Scientific Collaboration.

The LOSC has already released the first large dataset from Initial LIGO's flagship S5 run; more are planned to follow. The S5 release creates several concrete benefits.
It will promote broad participation in the advancement of GW physics and astrophysics from professional and amateur scientists, graduate students, undergraduates, and secondary students; all are invited to help verify and improve the quality of LIGO's scientific results.
The web services and tutorials created to enable GW research by scientists outside the LSC are already being used by researchers inside the collaboration, and are streamlining training and access to data for new LSC students.
The public dataset will provide excellent opportunities for outreach and education at many levels, and for many different audiences.
Finally, the effort to document and curate the S5 dataset has helped ensure that this archive will be a usable and valuable resource for many years to come.

LIGO will continue to release datasets through the LOSC, including data from the higher-sensitivity Advanced LIGO detectors that will begin operating in 2015 \cite{datamanagement}. Any GW signals identified by the LSC will be included in these releases. When regular GW detections begin to occur, public participation in LIGO data analysis will add an exciting dimension to GW astronomy. Numerical relativists, relativity theorists, astrophysicists, and others will use LIGO data to understand the dynamics of strongly curved spacetime, as well as the origins and properties of GW sources.

\section*{Acknowledgements}

We thank Stuart Anderson, Juan Barayoga, Patrick Brady, Duncan Brown, Laura Cadonati, Vuk Mandic, Szabi Marka, Greg Mendell, Keith Riles, Peter Shawhan, David Shoemaker, and Eric Thrane for helpful discussions and interactions.
Beverly Berger, Marie-Anne Bizouard, Vicki Kalogera, Andrew Lundgren, Vivien Raymond, and John Whelan participated in the LSC review of the S5 dataset.
We are grateful to the Virgo and GEO collaborations for allowing the LOSC Timeline tool to report observation-segment metadata for their detectors.
Alexander Cole, Ashley Disbrow, Gary LaMotte, and Shannon Wang performed early tests of the data. LIGO was constructed by the California Institute of Technology and Massachusetts Institute of Technology with funding from the National Science Foundation, and operates under cooperative agreement NSF-0757058. The LIGO Open Science Center (LOSC) was partially funded by grant NSF-1210172. Part of this work was carried out at the Jet Propulsion Laboratory, California Institute of Technology, under contract to the National Aeronautics and Space Administration.
This is LIGO document LIGO-P1400186. Copyright 2014.

\section*{References}

\bibliographystyle{iopart-num}

\bibliography{losc}

\end{document}